# Database of Wannier Tight-binding Hamiltonians using High-throughput Density Functional Theory

Kevin F. Garrity[1], Kamal Choudhary[1,2]

1 Materials Science and Engineering Division, National Institute of Standards and Technology, Gaithersburg, Maryland 20899, USA.

2 Theiss Research, La Jolla, CA, 92037, U.S.A.

**Abstract:**

We develop a computational workflow for high-throughput Wannierization of density functional theory (DFT) based electronic band structure calculations. We apply this workflow to 1771 materials, and we create a database with the resulting Wannier-function based tight binding Hamiltonians (WTBH). We evaluate the accuracy of the WTBHs by comparing the Wannier band structures to directly calculated DFT band structures on both the set of k-points used in the Wannierization as well as independent k-points from high symmetry lines. Accurate WTBH can be used for the calculation of many materials properties, and we include a few example applications. We also develop a web-app that can be used to predict electronic properties on-the-fly using WTBH from our database. The tools to generate the Hamiltonian and the database of the WTB parameters will be made publicly available through the websites https://github.com/usnistgov/jarvis and https://www.ctcms.nist.gov/jarviswtb.

**Background & summary**

Wannier functions (WF) were first introduced[1] in 1937, and have proven to be a powerful tool in the investigation of solid-state phenomenon such as polarization, topology, and magnetization[2]. Mathematically, WFs are a complete orthonormalized basis set that act as a bridge between a delocalized plane wave representation commonly used in electronic structure calculations and a localized atomic orbital basis that more naturally describes chemical bonds[3-5]. One of the most common ways of obtaining Wannier tight-bonding Hamiltonians (WTBH)[6-8] is by using the Wannier90 software package[9] to generate maximally localized Wannier functions, based on underlying density functional theory (DFT) calculations. However, obtaining high-quality Wannier functions requires several choices by code users, including which bands and energy ranges to Wannierize, as well as a choice of starting orbitals. Therefore, in order to unlock the many materials properties that can be calculated with WTBH for use in high-throughput computations, we provide tools to automate the Wannierization of DFT band structures, and we generate a database of verified WTBH for use in future applications.

The computational advantage of Wannier functions comes from their localization, which allows the WTBH to be determined once on a relatively coarse real-space grid, and then Fourier transformed to obtain the Hamiltonian and its derivatives at arbitrary k-points, allowing many expressions to be evaluated efficiently[10]. Many computationally expensive quantities such as $Z_2$ index, Chern number, Fermi-surfaces, Weyl-chirality, Hall conductivity, spin-texture, photo-galvanic effect, thermoelectric properties, thermal properties, Landau level applications, gyrotropic effects, and shift-photocurrent[9,11-16] can be efficiently computed with WTBHs. In addition, many materials properties are based on localized phenomena[17-19] such as impurities[20], defects[21,22], excitons[23], polarons[24], screened electron-electron interaction[25], and electron-phonon interactions[26], all of which can be modeled in a Wannier basis[27]. In addition, an examination of the Wannier Hamiltonian can provide an intuition to help understand bonding that is difficult to get from plane waves. They are also useful in second quantization based beyond DFT calculations such and Dynamical Mean Field Theory (DMFT) calulations[28,29].

The Materials Genome Initiative (MGI)[30] has recently spurred the generation of several high-throughput databases and tools such as from AFLOW[31], Materials-project[32], Open Quantum Materials Database (OQMD)[33], NOMAD[34], and NIST-JARVIS[35-46]. They have played key roles in the generation of electronic-property related databases for various material properties to foster industrial growth. However, the development of WTBH database and tools are still in the developing phase and require a thorough investigation [47-52].

The goal of this paper is to: a) develop a high throughput workflow for Wannierization of DFT calculations, b) develop a database of verified Wannier-based tight-binding Hamiltonians along with all related input/output files, c) develop web-apps for convenient WTBH predictions. We use our Wannierization workflow on the JARVIS-DFT (https://www.ctcms.nist.gov/~knc6/JVASP.html) database which is a part of the MGI at NIST. The NIST-JARVIS (https://jarvis.nist.gov/) has several components such as JARVIS-FF, JARVIS-DFT, JARVIS-ML, JARVIS-STM, JARVIS-Heterostructure and hosts material-properties such as lattice parameters[36], formation energies[38], 2D exfoliation energies[35], bandgaps[39], elastic constants[36], dielectric constants[39,53], infrared intensities[53], piezoelectric constants[53], thermoelectric properties[41], optoelectronic properties, solar-cell efficiencies[40,43], topological materials[38,39], and computational STM images[42]. The JARVIS-DFT database consists of ≈ 40000 3D and ≈1000 2D materials. As an initial step, we deploy our computational workflow on the materials that were recently predicted to be topologically non-trivial based on the spin-orbit spillage technique, including three dimensional (3D), two dimensional (2D), magnetic, non-magnetic, insulating, and metallic systems[38,39] systems including spin-orbit interactions. After obtaining the WTBH from DFT, we perform several checks to ensure the quality of the Hamiltonians. Although here we present results for mainly high-spillage materials, we will be extending this workflow to the entire JARVIS-DFT database. Currently, we have calculated Wannier Hamiltonians including spin-orbit coupling for 1406 3D and 365 2D materials, which can be used to efficiently calculate materials properties using either our software tools or other external software such as Wannier-tools[11], Z2Pack[54], WOPTIC[55], EPW[56]. We believe that releasing this database and toolset for use by the materials community should enable accelerated materials prediction and analysis.

**Methods**

The methodology supporting the current project consists of several steps that are given in Fig. 1.

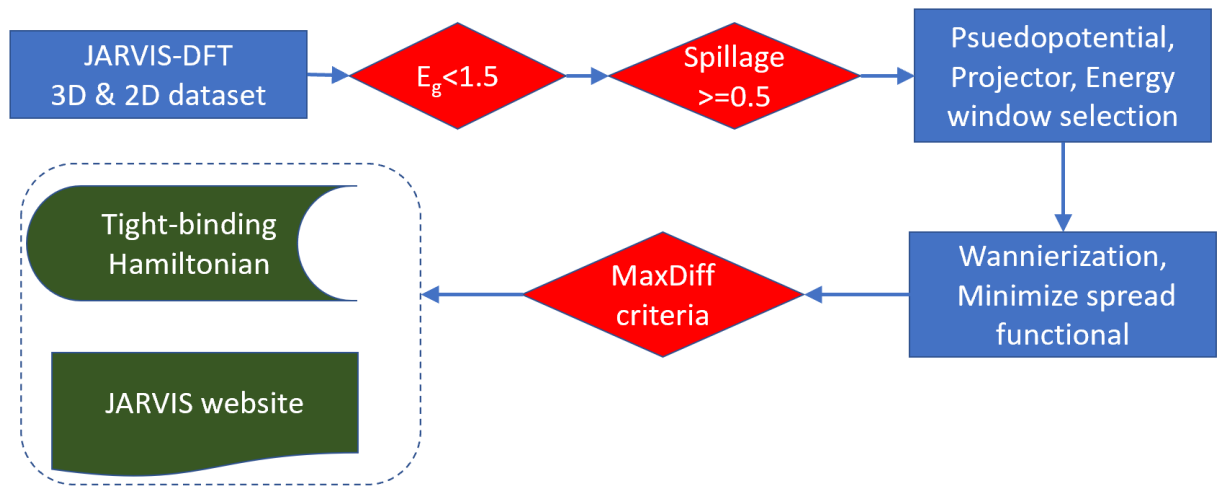

*Fig. 1 Workflow showing the Wannierization from using the DFT calculations.*

DFT calculations were carried out using the Vienna Ab-initio simulation package (VASP)[57,58] software using the workflow given on our JARVIS-Tools github page (https://github.com/usnistgov/jarvis). Please note commercial software is identified to specify procedures. Such identification does not imply recommendation by National Institute of Standards and Technology (NIST). We use the OptB88vdW functional[59], which gives accurate lattice parameters for both vdW and non-vdW (3D-bulk) solids[60]. We optimize the crystal-structures of the bulk and monolayer phases using VASP with OptB88vdW. Because spin-orbit coupling (SOC) is not currently implemented for OptB88vdW in VASP, we carry out spin-orbit PBE calculations. Such an approach has been validated by Refs. [38,61]. The crystal structure was optimized until the forces on the ions were less than 0.01 eV/Å and energy less than $10^{-6}$ eV. We use Wannier90[9] to construct Maximally-Localized Wannier Functions (MLWF) based TB-Hamiltonians.

The basic formalism of Wannierization is well-established. We briefly review some aspects here, interested readers can see longer discussions in [5,16]. For a set of Bloch eigenvectors $|\psi_{n,k}\rangle$, a general set of WFs $|\mathbf{R}n\rangle$ (n=1...N) can be written as:

$$|\mathbf{R}n\rangle = \frac{V}{(2\pi)^3} \int_{BZ} \sum_{m=1}^{N} U_{mn}^{(\mathbf{k})} |\psi_{mk}\rangle e^{-i\mathbf{k}\cdot\mathbf{R}} d\mathbf{k} \qquad (1)$$

where $\mathbf{R}$ labels the unit cell of the WF, $V$ is the volume of the unit cell, and $U_{mn}^{(\mathbf{k})}$ is an arbitrary unitary matrix. To construct maximally-localized WFs, $U_{mn}^{(\mathbf{k})}$ is chosen to minimize the following spread functional:

$$\Omega = \sum_n \left[ \langle r^2 \rangle_n - \bar{r}_n^2 \right] \qquad (2)$$

where $\bar{r}_n = \langle 0n|\mathbf{r}|0n\rangle$ and $\langle r^2 \rangle_n = \langle 0n|r^2|0n\rangle$. The minimization proceeds iteratively, based on an initial guess of localized orbitals.

For the case of interest in this work, where we wish to describe both the valence and conduction bands near the Fermi level, it is necessary to first select a set of bands to Wannierize, which includes separating the conduction bands from the free-electron-like bands that generally overlap with them in energy[62]. The procedure to determine this localized subspace of Bloch wavefunctions proceeds similarly to minimization described above, where after an initial guess, the subspace is iteratively updated in order to minimize the spread function in Eq. 2. After this initial disentanglement step, the Wannierization of the selected subspace proceeds as described above.

Due to the iterative non-linear minimization employed during both the disentanglement and Wannierization steps, the localization and utility of the final Wannier functions depend in practice on the initial choice of orbitals that are used to begin the disentanglement procedure, and which are then used as the initial guess for the Wannierization. Our initial guesses consist of a set of atomic orbitals we have chosen to describe all the chemically relevant orbitals for each element in typical elemental systems and compounds. We provide the list of the orbitals we select for each element in Table S1. For many specific materials, it may be possible to select a smaller set of orbitals while still maintaining high-quality WFs that describe the bands of interest; however, our fairly inclusive set of orbitals is able Wannierize nearly all compounds in a high-throughput manner without human intervention. Because most applications of WFs are computationally inexpensive compared to the DFT calculations used to construct the WFs, in practice, our larger Wannier basis has only minimal computational cost. However, it is necessary to have enough empty bands in the underlying DFT calculation such that any empty orbitals chosen are included in the Bloch basis. We do not include any semicore orbitals in our Wannier basis, as they are generally well-separated in energy from the valence orbitals and are not necessary to describe bands near the Fermi level, and we exclude semicore bands from the disentanglement.

During the disentanglement step, it is possible to choose an energy range that is included exactly ("the frozen window"), with the result that the Wannier band structure will exactly match the DFT band structure in this energy range and at the grid of k-points used in the Wannierization. We use a frozen window of ± 2 eV around the Fermi-energy. This window ensures that bands near the Fermi level are well described by the WTBH. For cases where the original WFs were unsatisfactory (see below), we found that lowering the lower bound of this window to include all of the valence bands often improves that WTBH, which we use as a second Wannierization setting. In order to validate our WTBH, we calculate the maximum absolute difference ($\mu$) between the Wannier and DFT eigenvalues within an energy range of ± 2eV around the Fermi level:

$$\mu = \max_{n\mathbf{k}} \left( \left| E_{n\mathbf{k}}^{DFT} - E_{n\mathbf{k}}^{WTB} \right| \right) \quad (3)$$

As discussed above, at the grid of k-points used in the construction of the WFs and within the frozen window, the eigenvalues should match exactly by construction. Therefore, to test the WTBH, we evaluate Eq. 3 on the dense lines of k-points used to generate our band structure plots, which tests the WFs out of sample. A weakness of this evaluation method is that highly dispersive energy bands (high $\frac{dE_{nk}}{dk}$) can result in high $\mu$ values even if the WTBH is of good quality because any slight shift in the *k*-direction of a dispersive band will result in a large energy error. We consider that systems with $\mu$ less than 0.1 eV to useful for most applications, and we provide data for the user to evaluate individual WTBH for their own applications.

**Data records**

After the calculations, the TB Hamiltonians, Wannier90 input and outputs files are stored as tar files which will be distributed through the JARVIS-DFT (https://www.ctcms.nist.gov/~knc6/JVASP.html)

website as well as the Figshare repository ( https://figshare.com/projects/JARVIS-DFT_Wannier_Tight-binding_Hamiltonians/82469). Each 'zip' file consists of wannier90.win, wannier90.wout, wannier90_hr.dat files. The wannier90.win and wannier90.wout are the input and output files for Wannier90 code respectively. The wannier90_hr.dat file can be loaded as WanHam class with scripts in the JARVIS-tools (https://github.com/usnistgov/jarvis) and similar packages to apply post-processing analysis such as calculating band-structures. There are also a JavaScript Object Notation (JSON) and Portable Network Graphic (PNG) file for comparing DFT bandstructure to WTBH.

**Technical validation**

To validate the WTBHs generated in this work, we compare the Wannier electronic bands with directly calculated DFT bands and measure the differences using Eq. 3 on two different k-point grids. As an example, in Fig. 1, we show an evaluation of the WTBH for $Bi_2Se_3$. In this figure, the top two panels show the WTBH evaluated on the same k-point grid used to generate the WFs, while the lower two panels show the evaluation on a typical set of high-symmetry k-points and lines, which includes k-points not used in the construction of the WFs. Figs. 1a and 1c show the eigenvalue comparison at separated k-points, with the WTBH bands in red and the DFT bands in blue, while Figure 1b and 1d show the eigenvalue differences as a function of energy.

As expected, the agreement within the frozen window and on the dense k-point grid is almost exact, but quickly increases up to 0.25 eV when leaving the window. We find a larger but still small energy difference on the high symmetry grid Fig. 1c-d, with a maximal error in the frozen window of 9 meV. This test shows that this WTBH can be used to interpolate the band structure accurately, and at the greatly reduced computational cost compared to DFT.

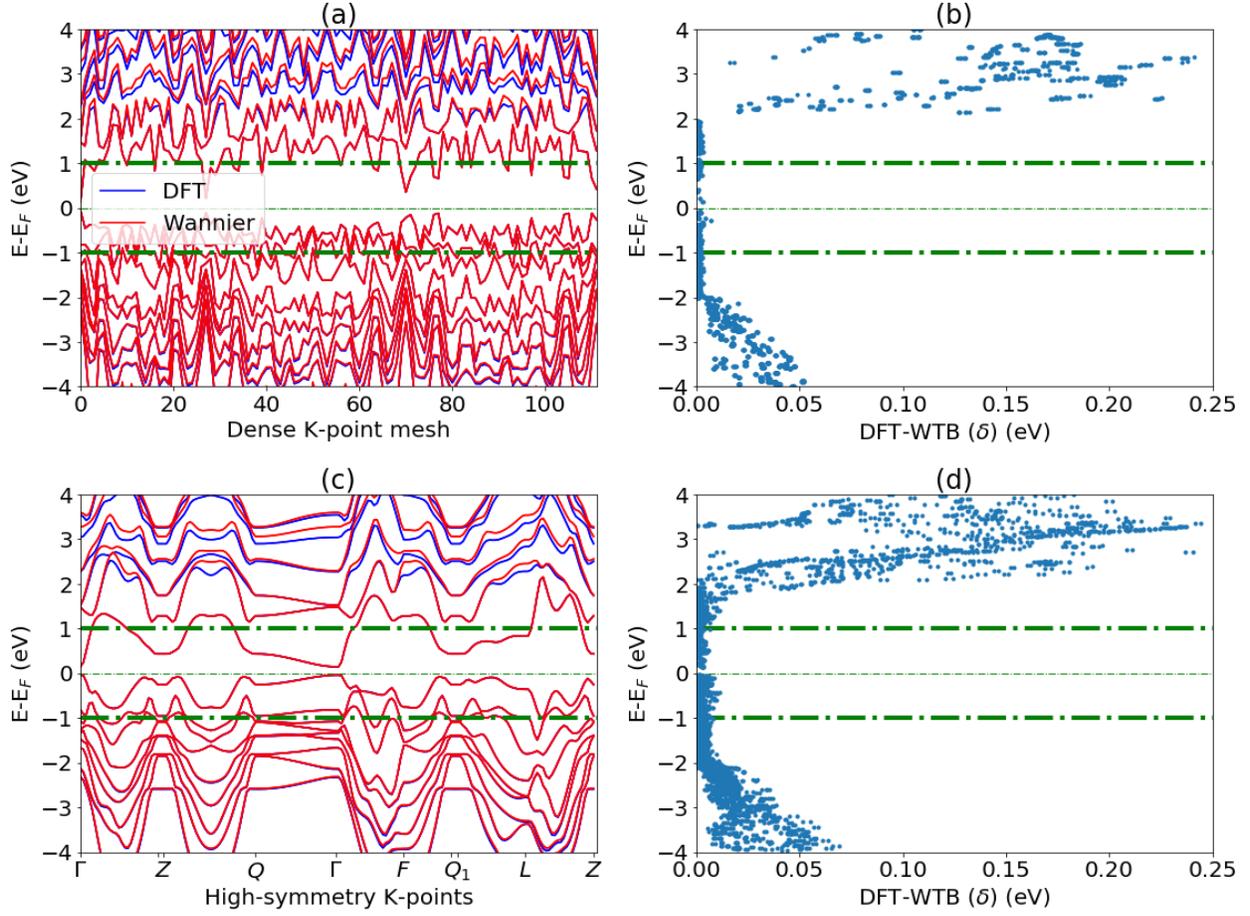

*Fig. 2 Comparison of DFT and WTB bandstructures for $Bi_2Se_3$. a-b) on dense k-grid, c-d) high-symmetry Brillouin zone points.*

We show a few more examples of 3D WTBH in Fig. 3 for Si, PbTe, $Sb_2Te_3$ and $Na_3Bi$, this time focusing only on the difference for the high-symmetry k-point grids. Similar to the $Bi_2Se_3$ case discussed above, they show the minimal difference, and the WTBH are able to reproduce features such as the Dirac point band crossing of $Na_3Bi$ between Γ and A.

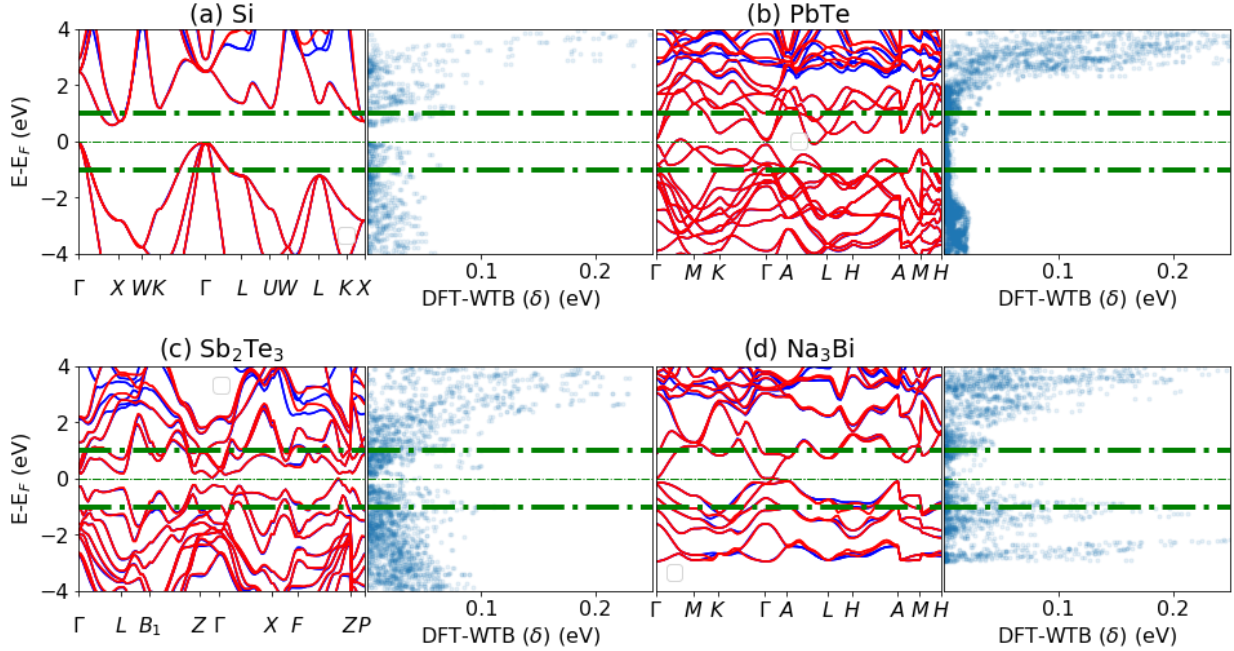

*Fig. 3 Examples of Wannier and DFT bandstructure and their energy difference plot for example 3D materials. a) Si, b) PbTe, c) Sb$_2$Te$_3$, and d) Na$_3$Bi.*

Bi$_2$Se$_3$, shown in Fig. 1, is a classic example of a 3D topological insulator. We show similar examples of 2D topological materials for graphene, ZrFeCl$_6$, Ti$_2$Te$_2$P, and VAg(PSe$_3$)$_2$ in Fig. 4. A detailed topological analysis of these materials can be found in our previous works[38]. Similar to the Bi$_2$Se$_3$ case, we observe that the DFT and WTBH bands overlap within the $\pm$ 2eV window and start to separate for outside these ranges. We again find excellent agreement between the DFT and the Wannier bands. Similar figures will be available for all the WTBH produced in this work on our website, so that the user can evaluate the WTBH for their own applications.

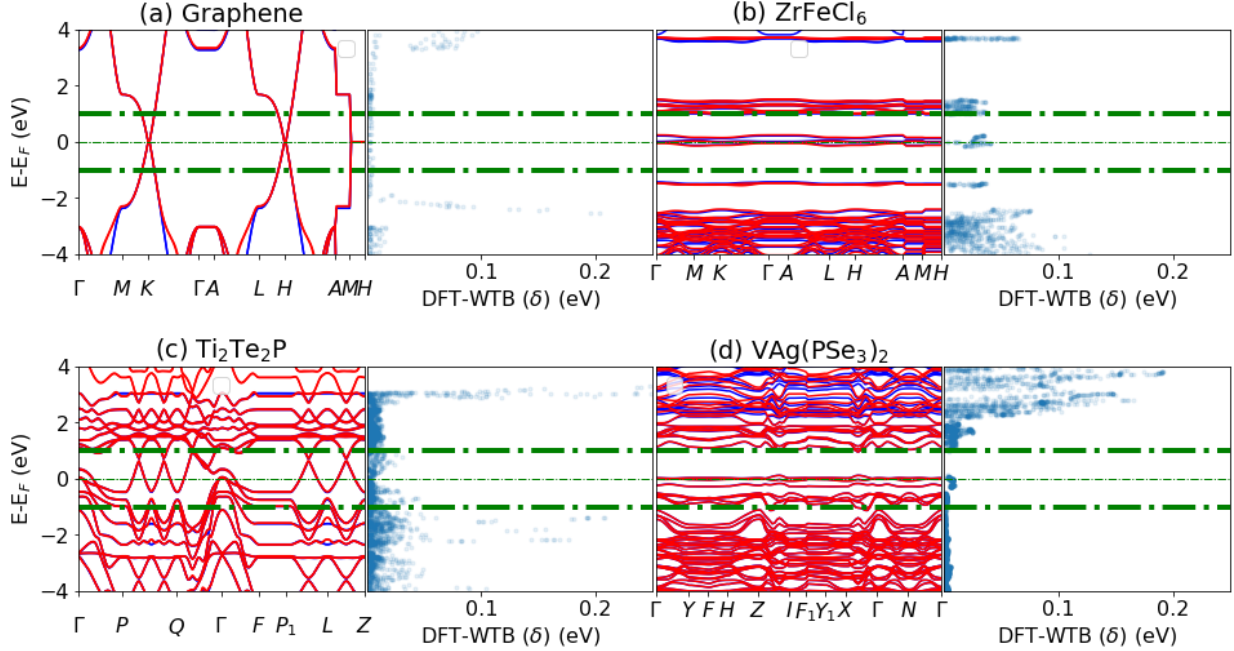

*Fig. 4 Examples of Wannier and DFT bandstructure and their energy difference plot for example 2D materials. a) for graphene, b) for ZrFeCl$_6$, c) for Ti$_2$Te$_2$P, and d) for VAg(PSe$_3$)$_2$.*

As is clear from the above examples, it is important to evaluate the energy difference between the DFT and WTBH bands to ensure a high-quality Wannierization. We use the maximum value of these differences (MaxDiff) for each k-point and in the disentanglement window range ($\pm 2$ eV) as the measure of the quality of WTBHs (see Eq. 6). We calculate these differences for both the k-point grid and high-symmetry BZ points. Choosing a tolerance of 0.1 eV as the maximum energy difference, we find that 93.0 % of materials have a dense k-mesh MaxDiff less than the tolerance, while only 64 % of materials have high-symmetry BZ MaxDiff less than the tolerance as shown in Fig. 5a and 5b respectively. These larger discrepancies mainly occur for metallic systems such as Al, which have very dispersive electronic bands that naturally result in larger errors as disused earlier (see also Fig. (S1)).

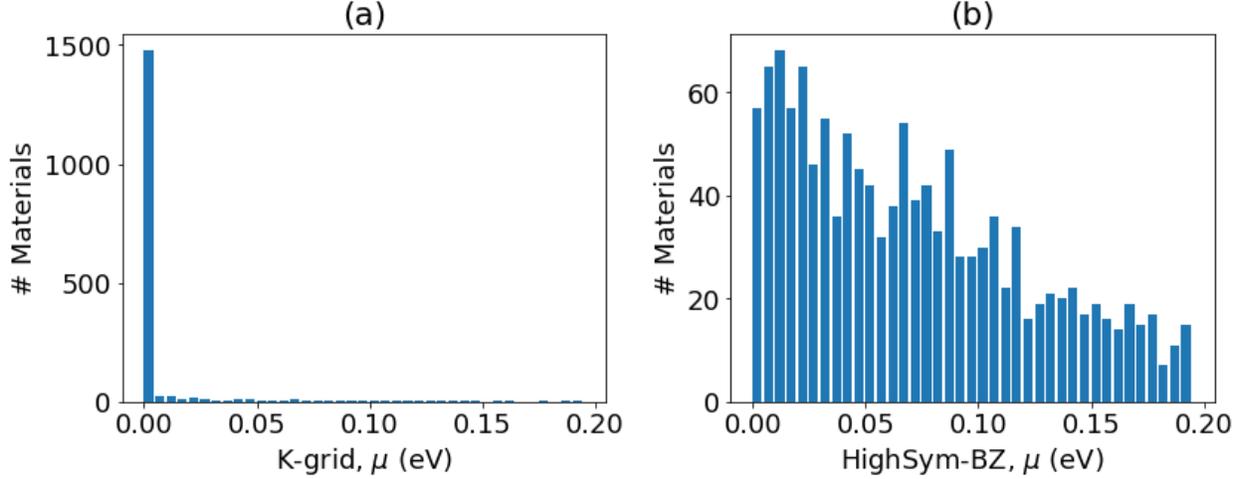

*Fig. 5 DFT-TB maximum difference (μ) distribution for all the Wannier Tight-binding Hamiltonians (WTBHs). A) on a regular k-point grid, b) on high-symmetry k-points.*

Next, we show a few example applications to demonstrate the usefulness of the WTB Hamiltonians. In Fig. 6a, we show the total and Bi (p) projected density of states in the $Bi_2Se_3$ system. The DOS can be evaluated with a very dense k-point grid at low computational cost using WFs, allowing detailed features to be converged. As mentioned in the introduction section, the WTB Hamiltonians can also be used to study defect phenomenon, especially if the defect only removes weak vdW bonds. For example, in Fig. 6b, we show the (001) surface bandstructure of $Bi_2Se_3$. As expected for a Z2 topological insulator, there is a bulk gap and a surface Dirac cone feature at Γ. Similarly, we show the edge band structure of a 2D monolayer of $VAg(PSe_3)_2$ with ferromagnetic spin ordering. $VAg(PSe_3)_2$ is a 2D Chern insulator[38], and the resulting spin-polarized conducing edge channel can be visualized in Fig.6c.

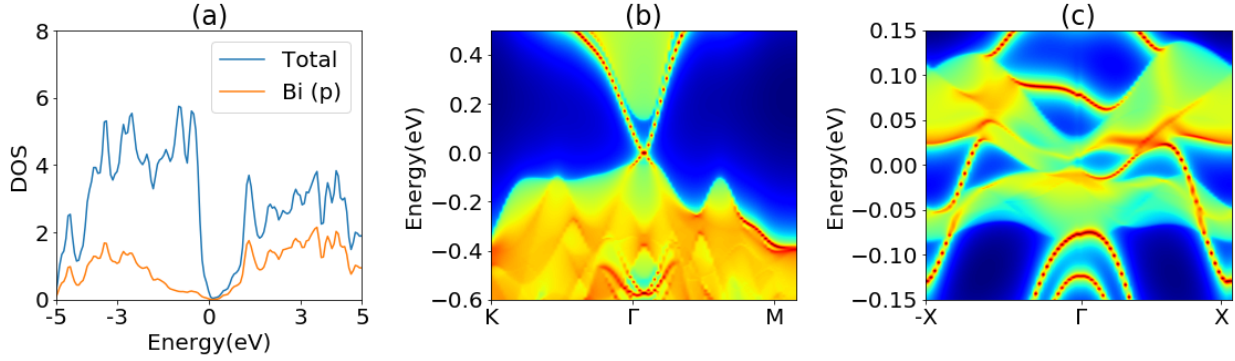

*Fig. 6 A few example applications of the WTB Hamiltonians. A) total and projected density of states, b) (001) surface band-structure of $Bi_2Se_3$, c) edge bandstructure of $VAg(PSe_3)_2$.*

Finally, in Fig. 7 we show a screenshot of a web-app we are developing to allow users to calculate materials properties using WTBH directly from our database, without downloading the Hamiltonians themselves. Currently, we support the calculation of Wannier-projected band structures for arbitrary k-points, as well as projected DOS. In addition, we provide plots to evaluate the accuracy of the WTBH. We plan to add other WTBH related functionalities in the app soon.

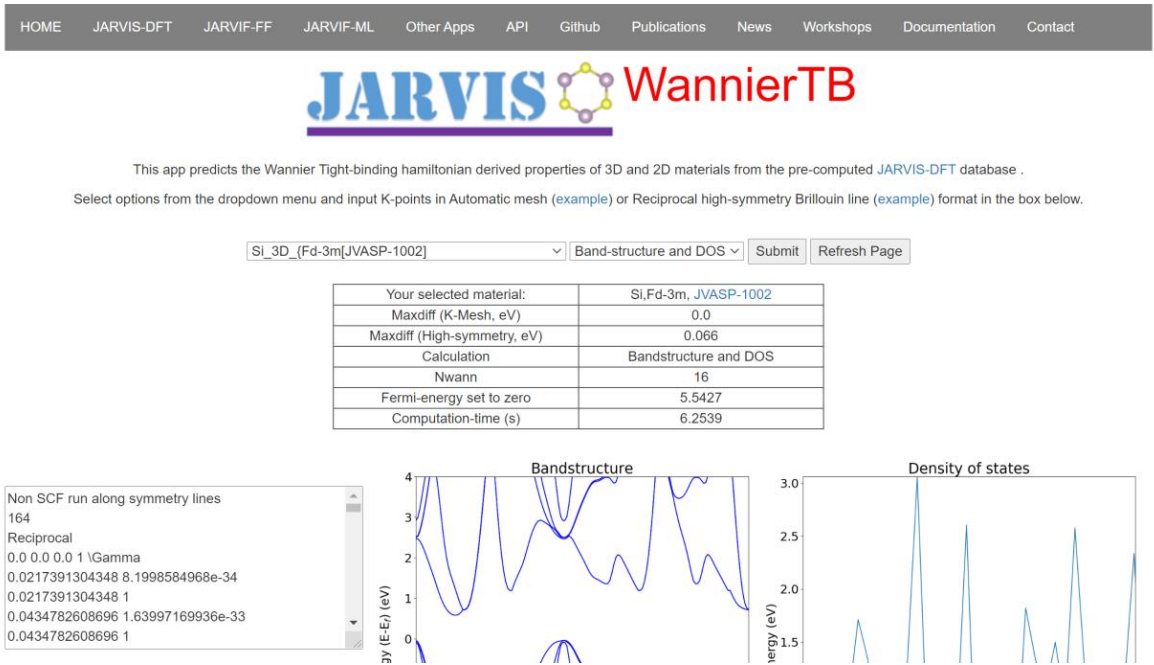

*Fig. 7 Snapshot of the web-app.*

**Usage notes**

The database presented here represents the largest collection of consistently calculated Wannier tight binding Hamiltonians of materials using density functional theory assembled to date. We anticipate that this dataset, and the methods provided for access will provide a useful tool in fundamental and application-related studies of materials. Our actual DFT verification provides insight into understanding the applicability and limitation of our the WTBH data. The WTBH can be used to obtain important electronic properties such as band-structures, density of states, and topological invariants in a computationally efficient way. Data-analytics tools can also be applied on the generated dataset.

**Contributions**

Both KFG and KC contributed in developing the workflow, analyzing data and writing the manuscript.

**Competing interests**

The authors declare no competing interests.

**Code availability**

Python-language based scripts for obtaining and analyzing the dataset are available at
https://github.com/usnistgov/jarvis

**Supplementary information: Database of Wannier Tight-binding Hamiltonians using High-throughput Density Functional Theory**

Kevin F. Garrity, Kamal Choudhary

Materials Science and Engineering Division, National Institute of Standards and Technology, Gaithersburg, Maryland 20899, USA.

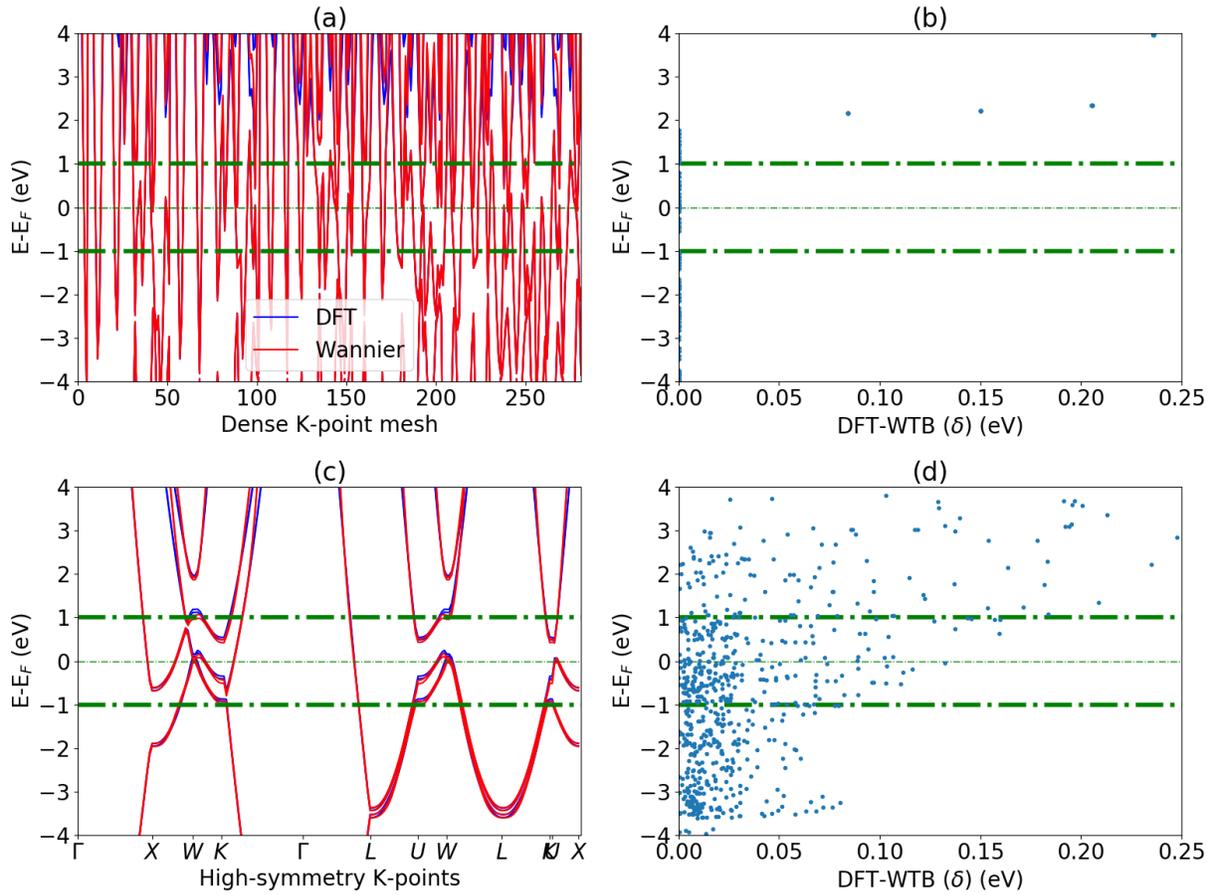

Fig. S1 *Comparison of density functional theory and Wannier tight-binding bandstructures for Al (JVASP-816). a-b) on dense k-grid, c-d) high-symmetry Brillouin zone points.*

Table S1: Semi-core states used during Wannierization. Note that several lanthanides and actinides are avoided.

| Element | POTCARs | N_electrons | Excluded_electrons | Projections | N_Wan_projections |
|---|---|---|---|---|---|
| **Ag** | Ag | 11 | 0 | s_d | 6 |
| **Al** | Al | 3 | 0 | s_p | 4 |
| **Ar** | Ar | 8 | 0 | s_p | 4 |
| **As** | As | 5 | 0 | s_p | 4 |
| **Au** | Au | 11 | 0 | s_d | 6 |
| **B** | B | 3 | 0 | s_p | 4 |
| **Ba** | Ba_sv | 2 | 8 | s_d | 6 |
| **Be** | Be_sv | 2 | 2 | s_p | 4 |
| **Bi** | Bi | 5 | 0 | s_p | 4 |
| **Br** | Br | 7 | 0 | s_p | 4 |
| **C** | C | 4 | 0 | s_p | 4 |
| **Ca** | Ca_sv | 2 | 8 | s_d | 6 |
| **Cd** | Cd | 2 | 10 | s_d | 6 |
| **Ce** | Ce | 4 | 8 | f_d_s | 13 |
| **Cl** | Cl | 5 | 2 | p | 3 |
| **Co** | Co | 9 | 0 | s_d | 6 |
| **Cr** | Cr_pv | 6 | 6 | s_d | 6 |
| **Cs** | Cs_sv | 1 | 8 | s_d | 6 |
| **Cu** | Cu_pv | 11 | 6 | s_d | 6 |
| **Dy** | Dy_3 | 12 | 3 | s_f | 8 |
| **Er** | Er_3 | 14 | 5 | f_s | 8 |
| **Eu** | Eu | 9 | 8 | f_s | 8 |
| **F** | F | 5 | 2 | p | 3 |
| **Fe** | Fe_pv | 8 | 6 | s_d | 6 |
| **Ga** | Ga_d | 3 | 10 | s_p | 4 |
| **Gd** | Gd | 10 | 8 | f_d_s | 8 |
| **Ge** | Ge_d | 4 | 10 | s_p | 4 |
| **H** | H | 1 | 0 | s | 1 |
| **He** | He | 2 | 0 | s | 1 |
| **Hf** | Hf_pv | 4 | 6 | s_d | 6 |
| **Hg** | Hg | 12 | 0 | s_p_d | 9 |
| **I** | I | 7 | 0 | s_p | 4 |
| **In** | In_d | 3 | 10 | s_p | 4 |
| **Ir** | Ir | 9 | 0 | s_d | 6 |
| **K** | K_sv | 1 | 8 | s_d | 6 |
| **Kr** | Kr | 8 | 0 | s_p | 4 |
| **La** | La | 3 | 8 | s_d_f | 13 |
| **Li** | Li_sv | 1 | 2 | s | 1 |
| **Lu** | Lu_3 | 17 | 8 | f_d_s | 13 |
| **Mg** | Mg_pv | 2 | 6 | s_p | 4 |
| **Mn** | Mn_pv | 7 | 6 | s_d | 6 |
| **Mo** | Mo_pv | 6 | 6 | s_d | 6 |
| **N** | N | 3 | 2 | p | 3 |
| **Na** | Na_pv | 1 | 6 | s_p | 4 |
| **Nb** | Nb_pv | 5 | 6 | s_d | 6 |
| **Nd** | Nd_3 | 6 | 5 | f_s | 8 |
| **Ne** | Ne | 8 | 0 | s_p | 4 |
| **Ni** | Ni_pv | 10 | 6 | s_d | 6 |
| **O** | O | 4 | 2 | p | 3 |
| **Os** | Os_pv | 8 | 6 | s_d | 6 |
| **P** | P | 5 | 0 | s_p | 4 |
| **Pb** | Pb_d | 4 | 10 | s_p | 4 |
| **Pd** | Pd | 10 | 0 | s_d | 6 |
| **Pt** | Pt | 10 | 0 | s_d | 6 |
| **Rb** | Rb_sv | 1 | 8 | s_d | 6 |
| **Re** | Re_pv | 7 | 6 | s_d | 6 |
| **Rh** | Rh_pv | 9 | 6 | s_d | 6 |
| **Ru** | Ru_pv | 8 | 6 | s_d | 6 |
| **S** | S | 4 | 2 | p | 3 |
| **Sb** | Sb | 5 | 0 | s_p | 4 |
| **Sc** | Sc_sv | 3 | 8 | s_d | 6 |
| **Se** | Se | 6 | 0 | s_p | 4 |
| **Si** | Si | 4 | 0 | s_p | 4 |
| **Sm** | Sm_3 | 8 | 3 | f_s | 8 |
| **Sn** | Sn_d | 4 | 10 | s_p | 4 |
| **Sr** | Sr_sv | 2 | 8 | s_d | 6 |
| **Ta** | Ta_pv | 5 | 6 | s_d | 6 |
| **Tb** | Tb_3 | 9 | 0 | f_s | 8 |
| **Tc** | Tc_pv | 7 | 6 | s_d | 6 |
| **Te** | Te | 6 | 0 | s_p | 4 |
| **Th** | Th | 4 | 8 | d_s | 6 |
| **Ti** | Ti_pv | 4 | 6 | s_d | 6 |
| **Tl** | Tl_d | 3 | 10 | s_p | 4 |
| **U** | U | 9 | 5 | f_s | 8 |
| **V** | V_pv | 5 | 6 | s_d | 6 |
| **W** | W_pv | 6 | 0 | s_d | 6 |
| **Xe** | Xe | 8 | 0 | s_p | 4 |
| **Y** | Y_sv | 3 | 8 | s_d | 6 |
| **Zn** | Zn | 12 | 0 | s_p_d | 9 |
| **Zr** | Zr_sv | 4 | 8 | s_d | 6 |